# Experimental demonstrations of unconditional security in a purely classical regime

Byoung S. Ham

Center for Photon Information Processing, School of Electrical Engineering and Computer Science,
Gwangju Institute of Science and Technology
123 Chumdangwagi-ro, Buk-gu, Gwangju 61005, S. Korea
(Submitted on August 14, 2020; bham@gist.ac.kr)

So far, unconditional security in key distribution processes has been confined to quantum key distribution (QKD) protocols based on the no-cloning theorem of nonorthogonal bases. Recently, a completely different approach, the unconditionally secured classical key distribution (USCKD), has been proposed for unconditional security in the purely classical regime. Unlike QKD, both classical channels and orthogonal bases are key ingredients in USCKD, where unconditional security is provided by deterministic randomness via path superposition-based reversible unitary transformations in a coupled Mach-Zehnder interferometer. Here, the first experimental demonstration of the USCKD protocol is presented.

**Introduction**

Quantum key distribution (QKD) has been intensively researched for unconditionally secured key distribution over the last several decades [1-13]. Since the first QKD protocol of BB84 [1], various types of QKD protocols have been successfully demonstrated using optical fibers, free space, and even satellites [10]. Regardless of QKD type, the essential requirement for unconditional security is lossless quantum channels and perfect single-photon detectors. Moreover, a deterministic nonclassical light source is required for potential applications of QKD such as online banking and quantum internet. So far, none of these requirements have been fully satisfied. As a result, the unconditional security of QKD lies in the no-cloning theorem based on Heisenberg's uncertainty principle [14] and cannot be fulfilled unless quantum loopholes are completely closed [6-13].

Recently, a completely different protocol for unconditionally secured classical key distribution (USCKD) has been proposed to overcome the limitations of QKD mentioned above as well as to understand the basic quantum features in a classical regime [15]. Compared with quantum superposition-caused randomness in QKD, USCKD achieves unconditional security via path superposition in a Mach-Zehnder interferometer (MZI), where a coupling method between two MZIs plays a key role [16]. Unlike QKD, USCKD is based on a purely classical system of MZIs with coherent light, where orthogonal phase bases are used. Thus, USCKD seems to be self-contradicting because the unconditional security of QKD is a quantum feature that cannot be obtained by classical means. Here, the quantum feature of unconditional security in USCKD is achieved via path superposition-caused measurement randomness for orthogonal bases, while the no-cloning theorem for QKD is provided by uncertainty-caused randomness for nonorthogonal bases. According to information theory, randomness represents that there is no information to eavesdrop [17]. Moreover, USCKD results in key distribution determinacy between two remote parties via the coherence physics of MZI, even without post-measurement of sifting, which is an essential step for QKD. The key distribution determinacy in USCKD is provided by reversible unitary transformations such as in quantum optical memories [18,19]. Thus, two-way communication channels are adapted to provide eavesdropping randomness and directional determinacy in a coupled MZI [15].

The fundamental physics of USCKD has been studied in a coupled MZI system [15], where a specific phase relationship between the coupled MZIs results in nonclassical features of coherence de Broglie waves (CBW) [16,20]. CBW is a classical version of photonic de Broglie waves (PBW), where PBW based on entangled photon pairs has been intensively studied for quantum sensing and quantum metrology over the last few decades [21-27]. Recently, experimental demonstrations of CBW have been successfully performed to demonstrate CBW theory, where USCKD represents the zero amplitude of CBW [28]. Thus, CBW has been understood as a macroscopic nonclassical feature [16]. In that sense, conventional understanding of the quantum nature limited to the microscopic world satisfying the uncertainty principle has been intrigued and expanded toward the macroscopic world, such as in the case of Schrodinger's cat [29,30]. Here, USCKD is experimentally studied for the proof of principle in a purely classical regime of a coupled MZI system. This study may open the door to coherence quantum technology, overcoming limitations in conventional quantum technologies confined to the microscopic world [1-13,21-27].

**Results**

Figure 1 shows an unfolded scheme of USCKD [15] based on orthogonal bases of coherent light for a classical key distribution, where two MZIs are coupled symmetrically with $\varphi_{12} = \psi_{12}$, where $\zeta_{ij} = \zeta_i - \zeta_j$. In each MZI, two separated paths are controlled by an AOM pair, in which each AOM driving frequency plays a key role for the phase control of the MZI. Here, the $\varphi_j$−based MZI belongs to Bob for key preparation, while the $\psi_j$−based MZI belongs to Alice to set the key. When Fig. 1 is folded for a round trip USCKD configuration, the right-end BS meets the left-end BS, which are then considered as identical. In other words, the detectors D3 and D4 with phase shifters B1 and B2 belong to Bob, while D1 and D2 with A1 and A2 belong to Alice. Each side has two available phase bases, $\varphi \in \{0, \pi\}$ and $\psi \in \{0, \pi\}$, respectively, where $\varphi \equiv \varphi_{12}$ and $\psi \equiv \psi_{12}$. To select a phase basis for each optical key, all four AOMs are synchronized to driving frequency generators, PTS160, PTS250, and AFG3102. The lower two AOMs, B2 and A2, are controlled by PTS160 and PTS250, respectively. The upper two AOMs, B1 and A1, are controlled by a two-channel AFG3102 (Tektronix). Thus, there are four possible phase basis combinations.

For the experiments, all AOMs are set to be in-phase, and the phase control of the MZI system relies only on the upper AOM A1 via AFG3102. For this, the lower AOM driving frequencies are kept at 80 MHz sharp. A sophisticated eavesdropper Eve can attack the transmission lines in both MZI channels, as shown by the red lines (e1 and e2) using a beam splitter pair. Such an attack is of course allowed in USCKD even without revealing her existence to Bob and Alice. Due to measurement randomness or indistinguishability in MZI, however, Eve's chance to extract the correct phase information is 50% on average, resulting in unconditional security [15]. This randomness of 50% is the bedrock of unconditional security in USCKD.

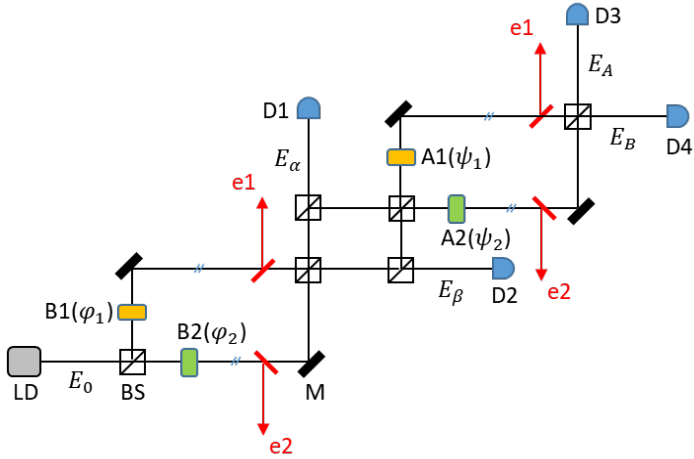

**Figure 1.** A schematic of unfolded USCKD. Aj($\psi_j$) and Bj($\varphi_j$) represent an acousto-optic modulator j for Alice and Bob with phase basis $\psi \in \{0, \pi\}$ and $\varphi \in \{0, \pi\}$, respectively, where $\psi = \psi_{12}(= \psi_1 - \psi_2)$ and $\varphi = \varphi_{12}(= \varphi_1 - \varphi_2)$. The e1 and e2 in red denote eavesdropping paths by Eve. LD: Laser, BS: unpolarizing beam splitter, M: mirror, and Dj: detector j. All Aj's are synchronized via microwave generators at 80 MHz.

*Theoretical approach*

For an analytic approach, the matrix representation for the first MZI in Fig. 1 is given:

$$\begin{bmatrix} E_\alpha \\ E_\beta \end{bmatrix} = [BS][\varphi][BS]\begin{bmatrix} E_0 \\ 0 \end{bmatrix} = \frac{1}{2}\begin{bmatrix} 1 - e^{i\varphi} & i(1 + e^{i\varphi}) \\ i(1 + e^{i\varphi}) & -(1 - e^{i\varphi}) \end{bmatrix}\begin{bmatrix} E_0 \\ 0 \end{bmatrix}, \quad (1)$$

where $\varphi = \varphi_{12}$ and $E_0$ is the input field of coherent light from LD. The BS matrix is $[BS] = \frac{1}{\sqrt{2}}\begin{bmatrix} 1 & i \\ i & 1 \end{bmatrix}$, and the matrix of a phase shifter between two MZI paths is $[\varphi] = \begin{bmatrix} 1 & 0 \\ 0 & e^{i\varphi} \end{bmatrix}$. Thus, the corresponding output intensities detected by D1 and D2 are as follows, respectively:

$$I_\alpha = \frac{1}{2}(1 - \cos\varphi), \quad (2)$$

$$I_\beta = \frac{1}{2}(1 + \cos\varphi), \quad (3)$$

where $I_j = E_j E_j^*$. Depending on the orthogonal phase basis of $\varphi \in \{0, \pi\}$ in MZI, the output field intensity becomes either $I_\alpha$ or $I_\beta$. Thus, Alice knows what basis is chosen by Bob by her measurements [15]. This represents the MZI propagation directionality. Here, it should be noted that the phase basis selection in $\varphi$ ($\psi$) belongs to Bob (Alice) for key preparation (confirmation) according to the USCKD protocol [15]. The output field from the first MZI is inserted into the second MZI for Alice's control. From the second MZI, the final output fields $E_A$ and $E_B$ are obtained as:

$$\begin{bmatrix} E_A \\ E_B \end{bmatrix} = [BS][\psi][BS]\begin{bmatrix} E_\alpha \\ E_\beta \end{bmatrix},$$

$$= -\frac{1}{2}\begin{bmatrix} e^{i\varphi} + e^{i\psi} & -i(e^{i\varphi} - e^{i\psi}) \\ i(e^{i\varphi} - e^{i\psi}) & e^{i\varphi} + e^{i\psi} \end{bmatrix}\begin{bmatrix} E_0 \\ 0 \end{bmatrix}. \quad (4)$$

Owing to the binary phase bases of $\varphi$ and $\psi$, there are four combinations of phase bases by Bob and Alice:

(i)      $\varphi = 0; \psi = 0$

For the case (i), equation (4) becomes:

$$\begin{bmatrix} E_A \\ E_B \end{bmatrix} = -e^{i\varphi}\begin{bmatrix} 1 & 0 \\ 0 & 1 \end{bmatrix}\begin{bmatrix} E_0 \\ 0 \end{bmatrix}. \quad (5)$$

Thus, the corresponding intensities are $I_A = I_0$ and $I_B = 0$.

(ii)      $\varphi = 0; \psi = \pi$

For the case (ii), equation (4) becomes:

$$\begin{bmatrix} E_A \\ E_B \end{bmatrix} = ie^{i\varphi}\begin{bmatrix} 0 & 1 \\ -1 & 0 \end{bmatrix}\begin{bmatrix} E_0 \\ 0 \end{bmatrix}. \quad (6)$$

Thus, the corresponding intensities are $I_A = 0$ and $I_B = I_0$.

(iii)      $\varphi = \pi; \psi = 0$

For the case (iii), equation (4) becomes:

$$\begin{bmatrix} E_A \\ E_B \end{bmatrix} = -ie^{i\varphi}\begin{bmatrix} 0 & 1 \\ -1 & 0 \end{bmatrix}\begin{bmatrix} E_0 \\ 0 \end{bmatrix}. \quad (7)$$

Thus, the corresponding intensities are $I_A = 0$ and $I_B = I_0$.

(iv)     $\varphi = \pi;\ \psi = \pi$

For the case (iv), equation (4) becomes:

$$\begin{bmatrix} E_A \\ E_B \end{bmatrix} = e^{i\varphi} \begin{bmatrix} 1 & 0 \\ 0 & 1 \end{bmatrix} \begin{bmatrix} E_0 \\ 0 \end{bmatrix}. \qquad (8)$$

Thus, the corresponding intensities are $I_A = I_0$ and $I_B = 0$.

In a short summary, $I_A = I_0$ (1) and $I_B = 0$ (0) are achieved for $\varphi = \psi$, otherwise $I_A = 0$ (0) and $I_B = I_0$ (1). Like Alice, Bob also knows Alice's phase basis choice by measuring his visibility even without communication with her. As a basic property of coherence optics, this propagation directionality is the quintessence of USCKD with superposition-caused measurement randomness [15].

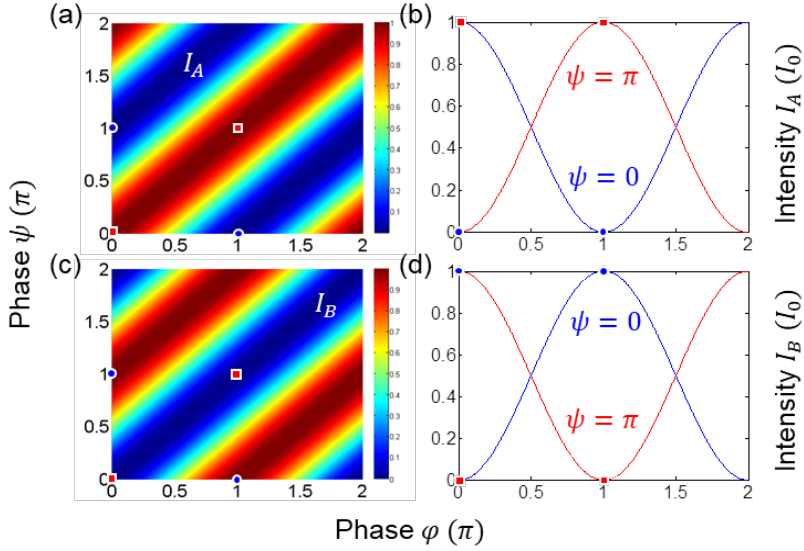

**Figure 2.** Numerical calculation for equation (4). $\varphi = \varphi_{12};\ \psi = \psi_{12}$. (a) and (b) $\varphi = \psi$. (c) and (d) $\varphi \neq \psi$. $\varphi, \psi \in \{0, \pi\}$. Red square (blue dot) indicates identity (inversion) relation between two phase bases.

These four options for the key distribution process analyzed in equations (5)-(8) are numerically demonstrated in Fig. 2 by solving equation (4). Figures 2(a) and (b) are for the output field $I_A$, and Figs. 2(c) and (d) are for $I_B$. Depending on the $\psi$−basis choice by Alice for a given $\varphi$−basis chosen by Bob, the output intensity becomes either $I_A$ or $I_B$. As an example, for $\varphi = \pi$ (see the red square marks), Alice confirms this with the same basis as shown in Figs. 2(a) and (b): $I_A = 1$ ($I_B = 0$) if $\psi = \varphi$; $I_A = 0$ ($I_B = 1$) if $\psi \neq \varphi$. The key distribution determinacy mechanism between Bob and Alice in USCKD is summarized in Table 1.

**Table 1.** Output fields in Fig. 1. $\varphi = \varphi_{12};\ \psi = \psi_{12}$. $I_j$ indicates 'on' or '1.'

|  |  | $\psi$ | |
|---|---|---|---|
|  |  | 0 | $\pi$ |
| $\varphi$ | 0 | $I_A$ | $I_B$ |
|  | $\pi$ | $I_B$ | $I_A$ |

*Experimental approach*

Figure 3 shows the experimental results corresponding to Fig. 2 and Table 1, where four different phase combinations are performed in a cw scheme of the laser light $E_0$. The temporal stability is determined mostly by air fluctuations in MZI paths. For Fig. 3, a rough laboratory condition is intentionally applied to the data without any system isolation or feedback phase control, where the MZI stability issue has already been closed [31,32]. Figure 3(a) shows the MZI channel stability for 20 seconds for the case of $\psi = \varphi \pm \pi$. For this, all four AOMs are set at 80 MHz and $\psi = \varphi$. Here, the experimental results of Fig. 3(a) are the same as in Fig. 2(b). As mentioned above, the experimental data are from bare laboratory conditions, resulting in ~20% phase (path length) fluctuations in short time scales less than a minute. In long-time scales, the output intensity varies between the minimum and maximum mostly due to air fluctuations.

Figure 3(b) shows a frequency-dependent phase control of the AOM A1. For this, the frequency for A1 is switched to either 1 Hz more or 1 Hz less than AOM A2 at 80 MHz sharp. The other AOMs are set at 80,000,001 Hz for B1 and 80,000,000 Hz for B2, resulting in $\varphi = 1\ Hz$ and $\psi = \pm 1\ Hz$. The asymmetric structure of the coupled MZIs for $\psi = -\varphi$ results in CBW, whose modulation frequency is doubled (2 Hz), as shown in the region left of the dashed line, because $\lambda_{CBW} = \lambda/2$ [16]. Here, the wavelength $\lambda$ is for the input light $E_0$, and $\lambda_{CBW}$ is due to the nonclassical properties of CBW. If $\psi = \varphi$ is satisfied, then the identity relation of equations (5) and (8) is satisfied (see both $I_A$ and $I_B$ in the right region of the dashed line), where the nonclassical feature of CBW disappears. The 1 Hz modulation is due to the background from the first MZI, which is not completely isolated in the experimental setup. Figure 3(c) is an extension of Fig. 3(b), where the output intensities of CBWs are also opposite each other as in the conventional MZI outputs in Fig. 2. Here, the zero modulation depth of CBW represents USCKD [33]. In other words, the modulation depth of CBW oscillates between two phase bases-determined extreme values, {0,1}, resulting in a half intensity at $\psi = \varphi + \frac{\pi}{2}$.

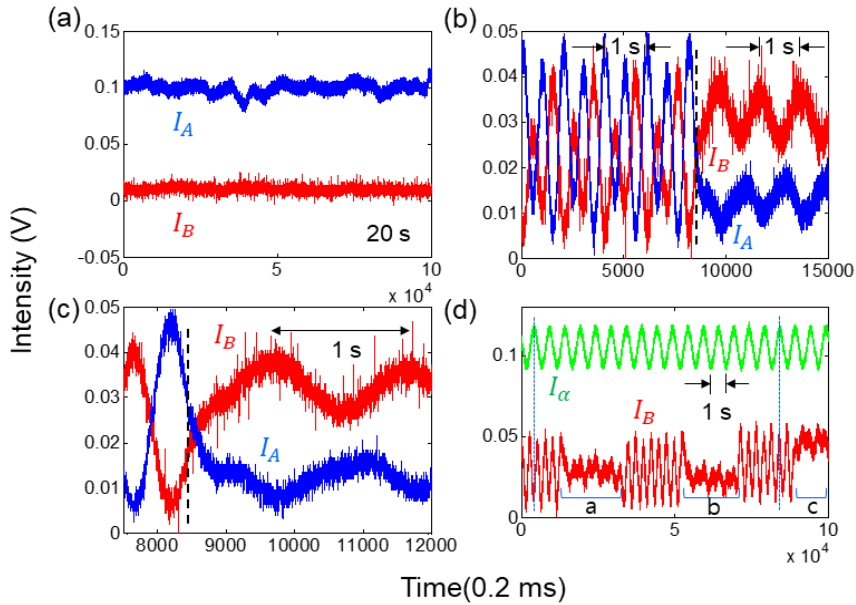

**Figure 3.** Experimental results for USCKD in Fig. 1. (a) $\psi_{12} = \varphi_{12} + \pi$ & $\psi_{12} = \varphi_{12} = 0$. (b) Switching between CBW and USCKD. (c) expansion of (b). (d) Conventional MZI output (green) vs. CBW [USCKD(a/b/c)] (red). In (b) and (c), $\varphi_{12} = -\psi_{12} = 1\ Hz$ before the dashed line; $\varphi_{12} = \psi_{12} = 1\ Hz$ after the dashed line. The value of vertical axis are arbitrary, where the photodetectors in Fig. 1 is controlled various attenuators.

Figure 3(d) represents toggle switching between CBW and USCKD for the reference of $I_\alpha$ of the first MZI. In the toggle switching process with AOM A2, the intensity value of USCKD depends on the phase of CBW at switching time as denoted in regions a, b, and c. For potential applications of USCKD, such an arbitrary intensity value can be controlled by controlling the internal phase of an rf generator. As already known for CBW bases [33], USCKD is understood as another form of CBW in terms of a symmetric mode in a coupled pendulum model [34]. The alternating CBW peaks between maxima and minima for a fixed value of $I_\alpha$ represent the increased phase bases (see the dotted lines). In other words, the π span in a single MZI is reduced to π/2 in the doubly coupled MZI. If an n-coupled MZI is used, then the phase basis span is reduced to π/n [33]. The related movie is shown in the Supplementary Information for toggle switching between CBW and USCKD.

Figure 4 shows snap shots of the output intensities from the oscilloscope for Fig. 3. Figure 4(a) corresponds to Fig. 3(b), where both the identity and inversion relations for USCKD in Fig. 2 are shown as results of toggle switching with AOM A2 from CBW. As shown, the maxima and minima of $I_A$ and $I_B$ are swapped according to a proper phase at the switching time. Figure 4(b) is for the reference $I_\alpha$ from the first MZI as a reference, whose modulation frequency is 1 Hz due to the preset 1 Hz driving frequency difference between AOMs B1 and B2. As mentioned in Fig. 2, Figure 4(c) shows $\psi$−dependent intensity swapping between $I_A$ and $I_B$, where the phase control is performed by manually rotating a thin glass inserted into the A1 path of Fig. 1 (see the bracket regions). Here, $\psi = \varphi = 0$ is set for Fig. 2(a). The rotation speed is not constant, but optimistically shows the trend of phase-dependent output-intensity variations. Glass rotation starts at a normal position with respect to the beam path, and thus the phase variation speeds up as it moves from region a to d.

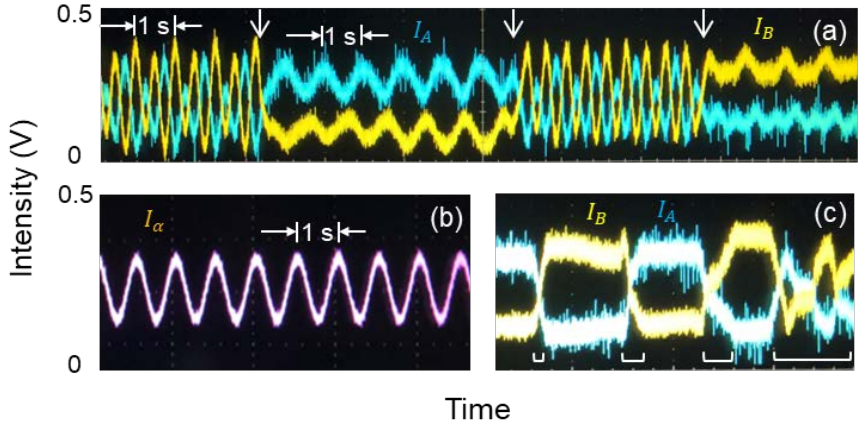

**Figure 4.** (a) CBW vs. USCKD. (b) Conventional MZI output $I_\alpha$. (c) Manual phase control for $\psi_1$ in Fig. 1. The driving frequencies for AOM B1 and B2 are 80,000,001 Hz and 80,000,000 Hz. The arrows indicate toggle switching timing. In (c), manual switching regions are denoted by a bracket.

**Discussion**

Unconditional security has been understood as an intrinsic property of QKD as a quantum feature, where experimental demonstration of USCKD should also provide the quantum features. As already discussed [15,16], such quantumness of USCKD lies in the coupled path superposition between two MZIs, which cannot be obtained classically. Thus, the coupled MZI structure should be differentiated from a single MZI, where the MZI belongs to the classical realm. To support the nonclassical property of the coupled MZIs, phase basis-based toggle switching with $\psi - \varphi = \pm 1$ Hz was demonstrated for swapping between CBW and USCKD. Here, the phase bases are orthogonal to each other, representing two modes of the nonclassical feature. All aspects of USCKD are macroscopic. Although the structure of MZIs for USCKD is definitely classical, coupled superposition results in nonclassical features of de Broglie waves in an asymmetric form and unitary transformations in a symmetric form. The unitary transformation represents deterministic randomness, where the superposition-caused randomness in MZI is the bedrock of unconditional security of USCKD [15]. Understanding that MZI is another form of BS, where orthogonal input modes are automatically provided [20], the nonclassical features of USCKD or CBW in the present demonstrations are not unexpected. Here, it should be noted that the physical origin of USCKD is the coupled superposition between two MZIs.

**Conclusion**

Experimental demonstrations of USCKD were presented in a symmetrically coupled MZI structure along with theoretical analyses. The unconditional security of USCKD was provided by deterministic randomness with round trip unitary transformations, where randomness plays a key role for unconditional security via MZI path superposition. The quantum behavior of the coupled MZI structure was confirmed by CBW with coupling manipulations, where the coupled MZIs regenerate fundamental phase bases. For the toggle switching between CBW and USCKD, a $\pm 1$ Hz frequency difference between the coupled MZIs was used. For the round trip MZI directionality of USCKD, a manual phase ($\psi$) variation with a thin glass was performed, where $0 \leq \psi \leq 2\pi$. The MZI stability was tested in bare conditions of MZIs without environmental isolations or a feedback control. Taking advantages of technologically advanced laser locking systems, an active control for MZI phase stability is not an issue anymore, and thus practical applications of USCKD are plausible for fiber-optic communication networks or free space in the future.

**Acknowledgments**
This work was supported by a GIST research institute (GRI) grant funded by GIST in 2020.

**Author contributions:** B.S.H. solely wrote the manuscript text and prepared all ideas, figures, calculations, and discussions. Correspondence and request of materials should be addressed to BSH (email: bham@gist.ac.kr)

**Competing interests:** The author declares no conflict of interest.

**Supplementary Information**

Supplementary information is available in the online version of the paper.